Thermal Relaxation Rates of Magnetic Nanoparticles in the Presence of Magnetic Fields and Spin-Transfer Effects

William Rippard, Ranko Heindl, Matthew Pufall, Stephen Russek, Anthony Kos

National Institute of Standards and Technology

Boulder, CO 80305

Abstract

We have measured the relaxation time of a thermally unstable ferromagnetic nanoparticle incorporated into a magnetic tunnel junction (MTJ) as a function of applied magnetic field, voltage $V$ (-0.38 V < $V$ < +0.26 V), and temperatures (283 K< T< 363 K) . By analyzing the results within the framework of a modified Néel-Brown formalism we determine the effective attempt time of the nanoparticle and also the bias dependences of the in-plane and out-of-plane spin transfer torques. There is a significant linear modification of the effective temperature with voltage due to the in-plane torque and a significant contribution of a "field like" torque that is quadratic with voltage. The methods presented here do not require complicated models for device heating or calibration procedures, but instead directly measure how temperature, field, and voltage influence the energy landscape and thermal fluctuations of a two-state system.  These results should have significant implications for designs of future nanometer-scale magnetic random access memory elements and provide a straightforward methodology to determine these parameters in other MTJ device structures.

Accurate and robust methods to determine fundamental parameters associated with nanoscale magnetic structures have been of significant interest within the magnetics community for a number of years. For both magnetic random access memory devices (MRAM) and the hard-disk drive industry, understanding thermal relaxation rates as well as the effects of the spin transfer torque[1-4] (STT) have become increasingly important as device and bit dimensions shrink. Individual magnetic elements must be thermally stable for data retention, while the writing process is generally strongly influenced by thermal fluctuations even at nanosecond switching times. Spin transfer effects are expected to enable future scaling of MRAM devices. However, there remains substantial uncertainty[5-17] in the functional dependence of the spin transfer torque on voltage. Here we demonstrate that we are able to quantitatively determine this dependence and the effective attempt time by measuring the thermal relaxation times (dwell times) of an individual superparamagnetic nanoparticle over a range of temperatures and as functions of applied field and voltage. This method is distinct from other room-temperature (RT) methods previously reported, and so provides an independent measurement of these parameters.

Numerous techniques have been used to determine the effective attempt time $\tau_0$ of magnetic nanoparticles including: single domain modeling at finite temperature[18], measurement of the relaxation rates at T ≈ 4 K[19,20], scanning tunneling microscopy studies[21], and modeling of the room-temperature field[22] or STT switching distributions[23], amongst others. The reported values for $\tau_0$ range over many orders of magnitude ($10^{-15}$ s to $10^{-9}$ s). While theoretical models predict some variation in $\tau_0$ due to its dependence on the magnetic properties of the device under study (*e.g.*, anisotropy, damping, and saturation magnetization)[24] and temperature, it is difficult to reconcile this wide range of values with these predictions. In addition, many of the methods are taken at parameter extremes (low temperature, high voltage), making the extrapolation back to the device operating point uncertain and model-dependent. Here we report on direct measurements of the relaxation times of magnetic nanoparticles at room temperature and low voltage, to determine the effective attempt time of nanoparticles that strongly resemble those relevant to MRAM devices and patterned bits for magnetic recording.

Similarly, many methods have been employed to determine the contributions of the "in-plane" and "field-like" torque terms associated with the STT effect including modeling the ferromagnetic resonance (FMR) frequency and linewidth changes associated with applied bias,[8-11] extraction from the size and symmetry of a RF current rectified by the MTJ itself,[12-15] determination from the phase relationship between the magnetization direction of a nanoparticle driven by an RF current,[16] and from the voltage dependent coercivity of an MTJ.[17] The reported results have ranged from the field like torque being completely linear with *V* to being completely quadratic in *V*. In addition, some models have ignored heating effects while the models that have attempted to include them have been forced to apply macroscopic heat transport theories to nanoscale objects. In the following, we quantitatively determine the in-plane and field-like STT dependencies on voltage from measurements of the relaxation times of magnetic nanoparticles at room temperature as a function of applied voltage. As we show below these results are, in general, robust with respect to model and heating effects, and identify where improvements in the modeling are possibly required.

The device studied here is etched from a MTJ stack consisting of Si substrate/SiO$_2$/TaCu(70)/PtMn(20)/CoFeB(2.5)/Ru(0.8)/CoFeB(2.5)/MgO(1.0)/CoFeB(2.5)/Ta(10)/Ru(5) (all thicknesses in nm) having a tunneling magnetoresistance TMR value of ≈ 110 % at room temperature. We pattern the free layer of the device into a nominally 65 nm x 90 nm ellipse through e-beam lithography and ion milling. The ion milling is stopped within (or just past) the MgO, leaving the pinned synthetic-antiferromagnetic layer thermally stable. From the wafer-level average (resistance•area)=5 Ω-μm$^2$ of the starting stack and the average device resistance of 950 Ω, we estimate that the actual device dimensions are 20 % larger than their nominal values, which is expected through fidelity loss in the pattern transfer via ion milling. The devices have uniaxial anisotropy predominantly due to magnetostatic shape anisotropy. For the 65 x 90 nm devices used here, the easy axis is along the device long axis with a typical anisotropy field of $\mu_0 H_k$ = 4 mT to 6 mT. As evidenced by the non-hysteretic TMR curve in Fig. 1(a), the device is thermally unstable, and retains the full TMR value. In all measurements reported below the applied field $H_{App}$ is applied along the nominal easy axis of the device.

The experimental measurement setup is shown in Fig. 1(b). A constant voltage is applied to a circuit in which the MTJ acts as variable resistor in series with a 1 KΩ resistor, with the voltage *V* across the MTJ being monitored through a buffer amplifier by a storage oscilloscope (bandwidth = 2.5 GHz). The buffer serves to isolate the relatively low impedance (50 Ω) scope from the high impedance MTJ. The practical bandwidth of the circuit is limited to roughly 50 MHz due to the RC time-constant associated with the high impedance MTJ and the capacitance of the cabling. The state of the device, parallel (P) or anti-parallel (AP), is monitored through the real-time voltage measured across the MTJ, an example of which is given in Fig 1b. As is shown in Fig. 1c, the voltage distribution is bi-modal with no intermediate states being present, and as such, the device is well-described as a bi-stable system. The relaxation time $\tau^{\pm}$ of the free layer magnetization at a particular applied field is determined by fitting $\exp(-t/\tau^{\pm})$ to a histogram of the times the device spends in a particular state (where $t$ is time and the superscript +(-) denotes the P to AP (AP to P) relaxation time, respectively) as shown in Fig 1d. Throughout this article, the histograms are created from a minimum of $10^3$ transitions, with a typical value being more than $10^4$. In Fig. 1, the voltage across the device is 6 mV (8 mV) in the P (AP) state, respectively. We consider this to be essentially equivalent to $V = 0$ V, the validity of which is verified below.

In the Néel-Brown model[24] the measured relaxation time (dwell time) of a magnetic particle with uniaxial anisotropy at $V = 0$ V as a function of applied field is given by:

$$\tau^{\pm} = \tau_0 \exp(\Delta(1 \pm \frac{H_{Eff}}{H_k})^n), \qquad (1)$$

where $\tau_0$ is the effective attempt time, $\Delta = E_0/k_B T$, where $E_0$ is the energy barrier, $k_B$ is Boltzmann's constant, $T$ is temperature, and $H_{Eff}$ is the total effective field applied along the easy axis of the device. In the following we take $n = 2$, which is appropriate for a single domain particle with uniaxial anisotropy.[25] We first analyze the relaxation times as a function of field and temperature (Fig. 2a) within the limit of $H_{Eff}/H_k \ll 1$ of Eq. (1):

$$\mathrm{Ln}(\tau^{\pm}) = \mathrm{Ln}(\tau_0) + \Delta(1 \pm \frac{2H_{Eff}}{H_k}). \qquad (2)$$

A fundamental prediction of Eq. (2) is that, assuming $\Delta$ for the P and AP states are equivalent, the relaxation times should be symmetric with respect to $H_{Eff}$, (*i.e.*, $|d\tau^{\pm}/dH_{Eff}|$ for the two states are equal). We use this as a criterion of device suitability, as it is not met by all structures. In many devices, symmetric behavior can be induced by rotating the device with respect to $H$, suggesting that the fabricated device is not parallel to the pin direction, although this was not necessary for the data presented here. In other cases, simple rotation does not strongly affect the asymmetric behavior, suggesting that it results from lithographic imperfections or intrinsic film defects. For all data shown here two separate time traces are recorded and analyzed under each bias condition in order to ensure repeatability of the measurements, with the resulting data points typically overlapping.

We can gain insight into the thermally activated nature of the switching by measuring the relaxation times as a function of temperature. Figure 2(a) shows the values of $\tau^{\pm}$ as a function of $H_{Eff}$ for $T = 283$ K and 363 K. Increasing temperature serves to reduce the relaxation times of both states, and also acts to change the coupling between the layers slightly. Figure 2(b) shows the ratio of the relaxation times $\mathrm{Ln}(\tau^{+}/\tau^{-})$ as a function of $H_{Eff}$ for several different temperatures, and shows that another effect of temperature is to change $d(\mathrm{Ln}(\tau^{+}/\tau^{-}))/dH_{Eff}$.

Within the NB model discussed above, at $\mu_0 H_{Eff} = 0$ T the relaxation times for the two states are equal, with a value $\tau^{Equal}$ given by

$$\text{Ln}(\tau^{Equal}) = \text{Ln}(\tau_0) + \frac{E_0}{k_B T}. \tag{3}$$

A plot of $\tau^{Equal}$ vs. $1/T$ is shown along with a fit in Fig. 2(c), which gives $\text{Ln}(\tau_0(s)) = -20 \pm 1.0$ (Ln(s)) (corresponding to $\tau_0 = 0.8$ ns to 6 ns) and $E_0 = (0.38 \pm 0.027)$ eV (corresponding to $\Delta(T=300\text{ K}) = 15 \pm 1.0$). This value of $\tau_0$ should be taken as a lower limit on its actual value, as it is possible that $E_0$ itself is temperature dependent, and a simple linear decrease in the value of $E_0$ with $T$ would serve to lower the apparent value of $\tau_0$, which cannot be accounted for in this measurement. Interestingly, we find that this room-temperature value of $\tau_0$ is on the same order as that reported for a nanoparticle at T = 4K.[19,20]

Varying the temperature also allows us to determine $H_k$ as well as estimate the effective switching volume of the device, which could be different from its physical volume if switching is being seeded by fluctuations in a subset of grains within it. In Fig. 2d we plot $d(\text{Ln}(\tau^+/\tau^-))/(d\mu_0 H_{Eff})$ as a function of $1/T$ along with a linear fit. From Eq (1), the slope of this function, *i.e.* $d^2(\text{Ln}(\tau^+/\tau^-)/d(1/T)d(\mu_0 H_{Eff})$, is $4E_0/(k_B\mu_0 H_k) = (3360 \pm 130)$ (K/mT), corresponding to $\mu_0 H_k = (5.2 \pm 0.18)$ mT using the value of $E_0$ determined above. This is in good agreement with the values from single domain modeling of $\mu_0 H_k \approx 4$ mT to 6 mT depending of the exact aspect ratio assumed for the actual device. Using the relation $E_0 = \mu_0 H_k M_s \Omega/2$, where $\Omega$ is the volume inducing switching, and the wafer level value of $M_s \approx 1000$ kA/m we find $\Omega = 2.3 \times 10^4$ nm$^3$, which is somewhat larger than the expected physical device volume of $1.3 \times 10^4$ nm$^3$, supporting the assertion that the device is switching as a single domain entity rather than via nucleation. In total, these data give strong evidence for the device acting as a single domain object being thermally activated over a single energy barrier, and that its switching is well-described by the NB model. The methodology and recent improvements in TMR values now allows such measurements to be performed at room temperature and low bias values, so that STT effects can be neglected.

We now turn our attention to the influence of voltage $V$ on the relaxation times as a function of $H_{Eff}$, shown for a range of $V$ (-0.38 V to 0.26 V) in Fig. 3a for $T$ = 303 K. For voltages outside this range the relaxation times were outside the bandwidth of the measurement circuit. Combined, these data represent the analysis of well over $10^6$ individual switching events. The relaxation times must be compared and analyzed at a constant voltage, as STT in MTJs is a voltage-driven process,[3,6] but for a given current bias the voltages across the P and AP states are significantly different. To account for this we first measure $\tau^-$ (the relaxation time for the AP to P transition) for a given current and then increase the current to induce the same voltage across the device while in the P state and then measure $\tau^+$. Hence, the values of $\tau^\pm$ for a given $V$ and $H_{Eff}$ shown in Figs 3 and 4 are taken from two different time traces.

The data in Fig. 3 show that the effects of $V$ are significant, which can be most easily seen by comparing the values of $\tau^\pm$ at $\mu_0 H_{Eff}$ = 0 mT. For $V$ = 8 mV, the relaxation times for $\mu_0 H_{Eff}$ = 0 mT are $\tau^\pm \approx$ 10 ms. For $V$ = +0.1 V, $\tau^-$ has been decreased by roughly a factor of 10, while $\tau^+$ has been increased by roughly the same factor, indicating that the voltage roughly stabilizes the P state and destabilizes the AP state equivalently. For $V = -0.1$ V the situation is very different. While $\tau^-$ has now been increased by a factor of 10, $\tau^+$ has been decreased by roughly a factor of 100. These variations are the combined effects of the in-plane and field-like spin transfer torques acting to alter the effective temperature and effective barrier between the two states, which are quantitatively analyzed below.

We analyze the relaxation times within a NB model modified to include the effects of the spin transfer torques[18, 26, 27]

$$\tau^\pm = \tau_0 \exp\left( \Delta \left(1 \pm \frac{V}{|V_{c0}^\pm|}\right) \left(1 \pm \frac{AV + BV^2 + H_{Eff}}{H_k}\right)^2 \right) \quad (4)$$

where $|V_{c0}^\pm|$ is the $T = 0$ K switching voltage. We allow for the field-like torque to have a linear and quadratic variation through the $(AV + BV^2)$ term while the in-plane torque serves to increase the effective temperature through the $(1 \pm V/|V_{c0}^\pm|)$ term.[26,28,29] We use the convention that $V$ is the voltage applied to the free layer while the fixed layer is held at ground.

In principle, we could directly fit the data in Fig. 3(a) with equation (4). However, we find that fitting the data in consecutive steps, as is done below, is more robust since it allows certain parameters to be removed from parts of the fitting process. The variation of $\text{Ln}(\tau^+/\tau^-)$ with $V$ and $H_{Eff}$ is given by:

$$\left(\frac{1}{\Delta}\right)\text{Ln}\left(\frac{\tau^+}{\tau^-}\right) = \left[4 + 2V\left(\frac{1}{|V_{c0}^+|} - \frac{1}{|V_{c0}^-|}\right)\right]\frac{H_{\text{Eff}}}{H_k} + \left[\frac{4A}{H_k} + \left(\frac{1}{|V_{c0}^+|} + \frac{1}{|V_{c0}^-|}\right)\right]V$$
$$+ \left[\frac{4B}{H_k} + \frac{2A}{H_k}\left(\frac{1}{|V_{c0}^+|} - \frac{1}{|V_{c0}^-|}\right)\right]V^2 + \left[\frac{2B}{H_k}\left(\frac{1}{|V_{c0}^+|} - \frac{1}{|V_{c0}^-|}\right)\right]V^3, \quad (5)$$

which is independent of $\tau_0$, simplifying the analysis. To arrive at Eq. (5) we have done a Taylor expansion of the quadratic term in Eq. (4), which is appropriate since $\text{Ln}(\tau^\pm)$ depends linearly on $H_{\text{Eff}}$ over the entire range of fields and voltages used here. Equation (5) predicts that $\text{Ln}(\tau^+/\tau^-)$ should vary linearly with $H_{\text{Eff}}$, with a slope that itself varies linearly with $V$, and have an intercept along the ordinate that could be a fairly complicated function of voltage depending on the relative strengths of the final three bracketed terms.

In obtaining Eq. 5 we have implicitly assumed that the value of $\Delta$, and hence $T$, for a given voltage is identical for the P and AP states even though the power dissipated across the device is different. At the highest voltages, for which heating is expected to be more significant, this approximation improves since the TMR ratio decreases and the resistances of the P and AP states converge (*e.g.,* the device TMR is 38% at $V = 0.3$ V). Similarly, we have implicitly assumed that for a given voltage the effective $\tau_0$ for the P and AP states are equivalent even though $\tau_0$ is predicted to be a function of the magnetic damping parameter and $T$.[24] The modification of $\Delta$ through the in-plane torque is expected to be much more important in describing $\tau^\pm$ than a potential modification of $\tau_0$ since $\Delta$ appears in the exponential term in Eq. (4). While there have been some attempts to account for Joule heating effects in MTJ structures[17,30], we have found that including them in our analysis does not yield robust results, but rather simply adds additional fitting parameters. In the following, the data are analyzed in such as way as to mitigate heating effects, and we show that the analysis is self-consistent over a significant voltage range and highlight where the inclusion of Joule heating in the model is possibly required.

The relaxation time ratios are shown in Fig. 3(b) along with linear fits, with their slopes given in Fig. 4a and intercepts along the field axis in Fig 4(b). As seen in Fig 3(b), the data are well-approximated by a linear relationship between $\text{Ln}(\tau^+/\tau^-)$ and $H_{\text{Eff}}$ over the entire range of $V$. As seen in Fig. 4(a), the slope $d(\text{Ln}(\tau^+/\tau^-))/dH_{\text{Eff}}$ varies linearly with voltage for $|V| < 200$ mV, and decreases outside this range. Restricting the analysis to the linear portion of the data, within the modified NB model the data should have a $V = 0$ V intercept of $4\Delta/\mu_0 H_k = 12$ (1/mT) and a slope $(d^2(\text{Ln}(\tau^+/\tau^-)/dH_{\text{Eff}}\, dV) = (2\Delta/\mu_0 H_k)$ $(1/|V_{c0}^+| - 1/|V_{c0}^-|) = -1.8$ (mTV)$^{-1}$, its negative value being consistent with $|V_{c0}^+| > |V_{c0}^-|$ found in spin torque switching experiments. Using these data we determine $(1/|V_{c0}^+| - 1/|V_{c0}^-|) = -0.31$ (1/V), which is used below. The decrease in $d(\text{Ln}(\tau^+/\tau^-))/dH_{\text{Eff}}$ at the higher voltages is consistent with device heating (which would directly lower $\Delta$), a sub-linear increase of the in-plane torque at large $V$ as was measured in Ref [16], or a combination of the two. More detailed measurements are needed to distinguish these potential contributions.

Figure 4(b) shows the measured values for the field intercept of $\text{Ln}(\tau^+/\tau^-)$ as a function of voltage, along with a fit to a quadratic function. Using the measured value $(1/|V_{c0}^+| - 1/$ $|V_{c0}^-|$) = - 0.31 (1/V) and the approximation that $2 \gg |V|(1/|V_{c0}^+|-1/|V_{c0}^-|)$, which is valid over the voltage range studied here, these data are described as:

$$H(\tau^+ = \tau^-) = -\left\{\left[A + \frac{H_k}{4}\left(\frac{1}{|V_{c0}^+|} + \frac{1}{|V_{c0}^-|}\right)\right]V + \left[B + \frac{A}{2}\left(\frac{1}{|V_{c0}^+|} - \frac{1}{|V_{c0}^-|}\right)\right]V^2\right\}. \quad (6)$$

It is important to note that $\Delta$ does not appear in this equation (within the approximation that $\Delta$ for the P and AP states is equal under equal $V$). Hence, in this analysis any voltage dependent effects of Joule heating that act to lower $\Delta$ are minimized.

The data in Fig 4(b) have significant linear and quadratic components. From room temperature measurements of thermally stable devices having a similar area but elongated shape (50x150 nm$^2$) from the same wafer, we estimate that $|V_{c0}^+| \approx 0.9$ V and $|V_{c0}^-| \approx 0.7$ V under low bias conditions.[31] Hence, the linear term evident in Fig. 4(b) largely arises from the modification of $\Delta$ from the in-plane torque, since $(\mu_0 H_k/4)(1/|V_{c0}^+| + 1/|V_{c0}^-|) \approx 3.3$ mT/V. From this, we can estimate the value of $A \approx 1.1$ mT/V, which is substantially smaller than reported in Ref. [9] but similar to Ref. [11]. The sign of $A$ indicates that, in this sample, it acts to stabilize the P (AP) state for $V > 0$ V ($< 0$ V), respectively. Using this value for $A$ and our determined value for $(1/|V_{c0}^+|-1/|V_{c0}^-|)$, the quadratic term is dominated by the contribution of $B = (-3.2 \pm 0.16)$ mT/V$^2$, consistent with Refs [14-16] but significantly less than in Refs [8,10,32]. We note that while this analysis should be relatively insensitive to heating effects mentioned in regard to Fig. 4(a), a saturation in-plane torque would serve to increase the value of $|B|$. Using these determined values of $A$ and $B$, we can conclude that the 6 mV to 8 mV bias used in deriving values for $\Delta$, $\tau_0$, and $H_k$ above had no significant effect on their values.

The sign of *B* indicates that it favors AP alignment within the device independent of the sign of *V*, in agreement with Refs [8,14-17]. Over the voltages studied here, we see no indication of the field like torque deviating from its quadratic dependence, unlike in Ref [16]. The fact that the quadratic term favors AP alignment in the device is a positive feature for STT-RAM devices in the low voltage regime, since it is for P to AP switching that the required voltages are largest. However, in the large voltage regime it can also be responsible for "back hopping".[17,33]

In summary, we have measured the relaxation times of a thermally unstable ferromagnetic nanoparticle as functions of temperature, field, and voltage. From low bias measurements, the "intrinsic" properties of the device $\tau_0$, $\Delta$, and $H_k$ can be determined. The effect of an applied voltage is to modify the relaxation times of the two states through both in-plane and field-like torques, which change the effective temperature and barrier height respectively. By analyzing the results with a modified NB model, we are able to quantitatively determine the quadratic dependence of the field-like torque and estimate a linear contribution. The methods presented here should be easily applicable to other materials systems and devices, allowing systematic comparison between them.

Acknowledgements:
We thank Everspin Technologies for providing the starting MTJ material, W. Butler and M. Stiles for helpful discussion regarding the preparation of this manuscript, and M. Keller for help with the data analysis software. This work was partially supported by the DARPA STT-RAM program.

References:

[1]  J. Slonczewski, J. Magn. Magn. Mat. **159**, L1-L7 (1996).

[2]  J. Slonczewski.and J. Sun, J. Magn. Magn. Mat. **310**, 169-175 (2007).

[3]  J. Slonczewski,, Phys. Rev. B **71**, 024411 (2005).

[4]  L. Berger, Phys. Rev. B **54**, 9353-9358 (1996).

[5]  A. Kalitsov, M. Chshiev, I. Theodonis, N. Kioussis, and W. Butler, Phys. Rev. B **79**, 174416 (2009).

[6]  C. Heiliger and M. D. Stiles, Phys. Rev. Lett. **100** 186805 (2008).

[7]  J. Xiao, G. E. W. Bauer, and A. Brataas, Phys. Rev. B **77**, 224419 (2008).

[8]  A.M. Deac, A. Fukushima, H. Kubota, H. Maehara, Y. Suzuki, S.Yuasa, Y. Nagamine, K.Tsunekawa, D. D. Djayaprawira, and N. Watanabe, Nat. Phys. **4**, 803-809 (2008).

[9]  O. Heinonen, S. Stokes, and J. Yi, Phys. Rev. Lett. **105**, 066602 (2010).

[10]  M. H. Jung, S. Park, C.-Y. You, and S. Yuasa, Phys. Rev. B **81**, 134419 (2010).

[11]  S. Petit, C. Baraduc, C. Thirion, U. Ebels, Y. Liu, M. Li, P. Wang, and B. Dieny, Phys. Rev. Lett. **98**, 077203 (2007).

[12]  J. C. Sankey, P. M. Braganca, A. G. F. Garcia, I. N. Krivorotov, R. A. Buhrman, and D. C. Ralph, Phys. Rev. Lett. **96**, 227601 (2006).

[13]  A. A. Tulapurkar, Y. Suzuki, A. Fukushima, H. Kubota, H. Maehara, K. Tsunekawa, D. D. Djayaprawira, N. Watanabe, and S. Yuasa, Nature **438**, 339-42 (2005).

[14]  Kubota, H. et al., Nat. Phys. **4**, 37-41 (2007).

[15]  C. Wang, Y.-T. Cui, J. Sun, J. A. Katine, R. A. Buhrman, and D. C. Ralph, Phys. Rev. B **79**, 224416 (2009).

[16]  C. Wang, Y. -T. Cui, J. A. Katine, R. A. Buhrman, and D. C. Ralph, Nat. Phys. **7**, 496 (2011).

[17]  S.-C Oh et al., Nat. Phys. **5**, 898-902 (2009).

[18]  H-J Suh, C. Heo, C.-Y. You, W. Kim, T.-D. Lee, and K.-J. Lee, Phys. Rev. B **78**, 064430 (2008).

[19] W. Wernsdorfer, E. Orozco, K. Hasselbach, A. Benoit, B. Barbara, N. Demoncy, A. Loiseau, H. Pascard, and D. Mailly, Phys. Rev. Lett. **78**, 1791-1794 (1997).

[20] I. N. Krivorotov, N. C. Emley, A. G. F. Garcia, J. C. Sankey, S. I. Kiselev, D. C. Ralph, and R. A. Buhrman, Phys. Rev. Lett. **93**, 166603 (2004).

[21] S. Krause, G. Herzog, T. Stapelfeldt, L. Berbil-Bautista, M. Bode, E. Vedmedenko, and R. Wiesendanger, Phys. Rev. Lett. **103**, 127202 (2009).

[22] C. P. Hancock and K. O'Grady, and M. El-Hilo, .J. of Phys. D: App. Phys. **29**, 2343–2351 (1996).

[23] D. Bedau, H. Liu, J. Z. Sun, J. A. Katine, E. E. Fullerton, S. Mangin, and A. D. Kent, App. Phys. Lett. **97**, 262502 (2010).

[24] W. F. Brown, Jr.. Phys. Rev. **130**, 1677-1686 (1963).

[25] R. H. Victora, Phys. Rev.Lett. **63,** 457-460 (1989).

[26] Z. Li and S. Zhang, Phys. Rev. B **69**, 134416 (2004).

[27] Z. Li, S. Zhang, Z. Diao, Y. Ding, X. Tang, D. Apalkov, Z. Yang, K. Kawabata, and Y. Huai, Phys. Rev. Lett. **100**, 246602 (2008).

[28] R. H. Koch, J. A. Katine, and J. Z. Sun, Phys. Rev. Lett. **92**, 088302 (2004).

[29] J. Z. Sun, Phys. Rev. B **62**, 570-578 (2001).

[30] G. D. Fuchs, I. N. Krivorotov, P. M. Braganca, N. C. Emley, A. G. F. Garcia, D. C. Ralph, and R. A. Buhrman, App. Phys. Lett. **86**, 152509 (2005).

[31] R. Heindl, W. H. Rippard, S. E. Russek, M. R. Pufall, and A. B. Kos, J. App. Phys. **109**, 073910 (2011).

[32] S. Petit, N. de Mestier, C. Baraduc, C. Thirion, Y. Liu, M. Li, P. Wang, and B. Dieny, Phys. Rev. B **78**, 184420 (2008).

[33] J. Z. Sun, M. C. Gaidis, G. Hu, E. J. O'Sullivan, S. L. Brown, J. J. Nowak, P. L. Trouilloud, and D. C. Worledge, J. App. Phys. **105**, 07D109 (2009).

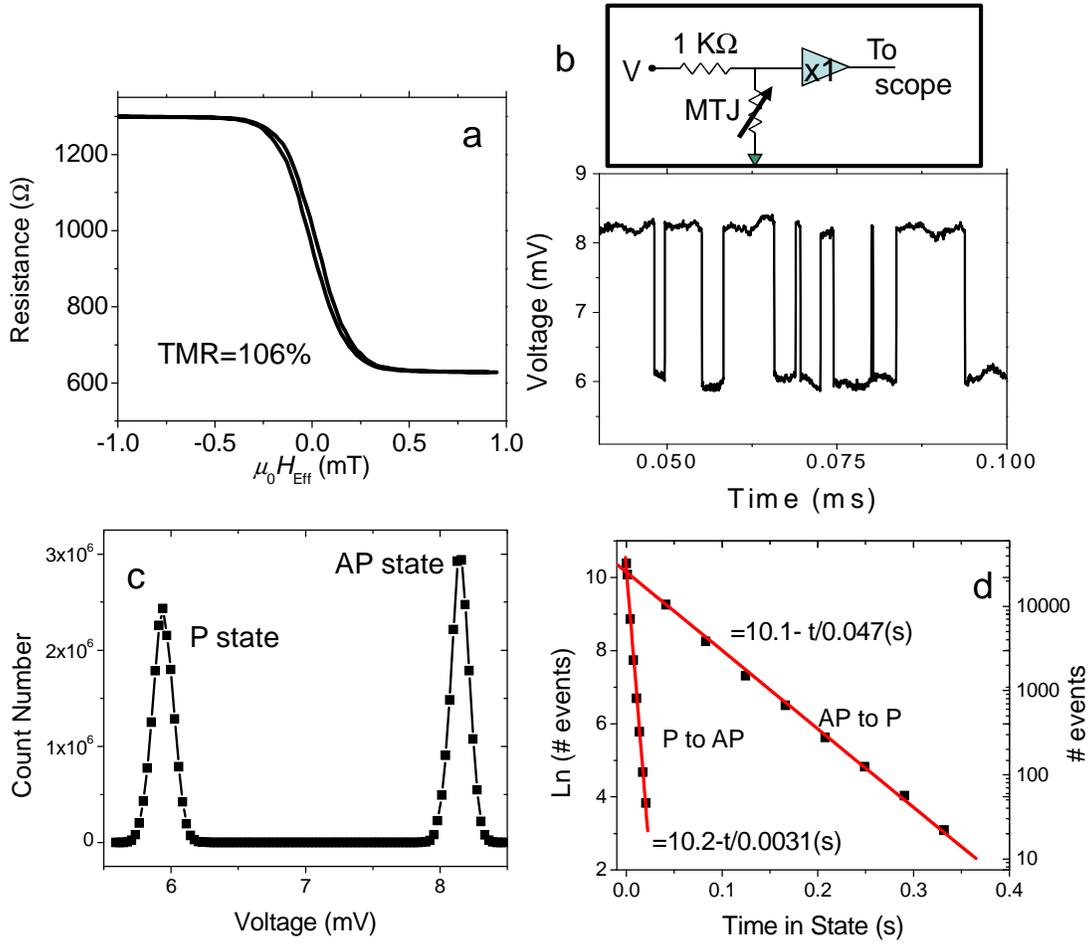

FIG 1 (a) Experimental TMR curve for the device shown here taken with a 5 µA current bias. The coupling field $H_{coup}$ (= -0.9 mT at T = 303 K ) between the CoFeB layers has been subtracted, as it is in all subsequent data. In all following data we define $H_{Eff}=(H_{applied}-H_{coup})$. (b) A small section of an experimental time trace showing two state switching for $\mu_0 H_{Eff}$= 0 mT. The voltage across the device corresponds to $V^{AP}$=8 mV and $V^{p}$=6 mV. The device is incorporated into a voltage divider circuit, so the difference in the voltages for the P and AP states does not correspond to the TMR value of the device. The full time trace consists of 32 Mpts and captures a roughly $4\times10^4$ transitions. (measurement schematic) The sampling rate of the scope is set so that a high number of transitions is obtained while the sampling rate is a minimum of 200 times faster than the relaxation rate, so that short dwell time events are not missed. (c) A histogram of the full time trace of the measured voltage across the device shown in (b). The bi-modal structure indicates that the device is well-described by simple two-state switching with no evidence of an intermediate state being present. (d) Representative data showing the histogram of the time ($\mu_0 H_{Eff}$ = -0.2 mT) the device remains in the P and AP states. The fits to the data give $\tau^{\pm}$ through $Ln(N) \sim -t/\tau^{\pm}$ where $N$ is the number of events.

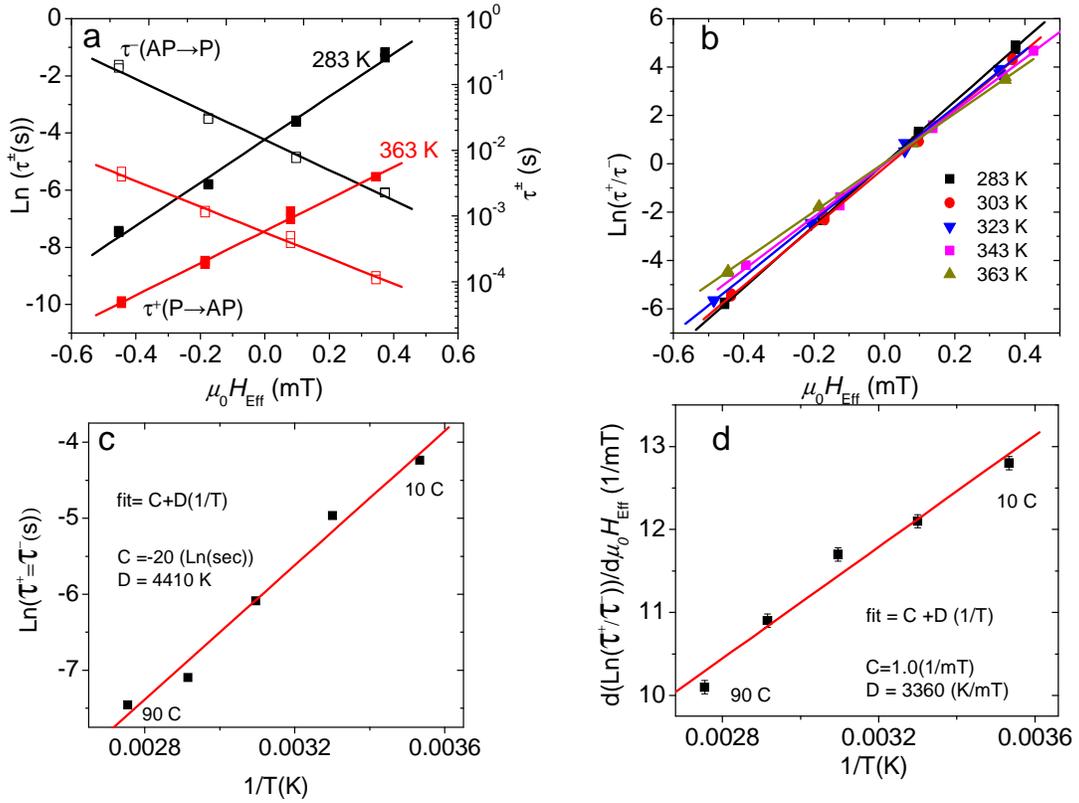

FIG 2 (a) Measured relaxation times for both $\tau^+$ (the P-to-AP (■) transition) and $\tau^-$ ( the AP-to-P (□) transition) as functions of field for $T$= 283 K and 363 K (other temperatures are not shown for clarity). (b) Ratio of the relaxation times as a function of field for $T$ ranging from 283 K to 363 K along with linear fits. We define $\mu_0 H_{Eff}$ = 0 mT as the field for which $\tau^+=\tau^-$. The value of $\mu_0 H_{coup}$ changes slightly with $T$ ($\approx$ 0.002 mT/K) requiring different values of $H_{Coup}$ to be subtracted from $H_{App}$ for each temperature. (c) The values of $\tau^{Equal}$ vs $1/T$. The slope of this line gives $E_0$ and its intercept $Ln(\tau_0)$. (d) $d(Ln(\tau^+/\tau^-))/d(\mu_0 H_{Eff})$ vs $1/T$, the error bars are determined from the fits in (b).

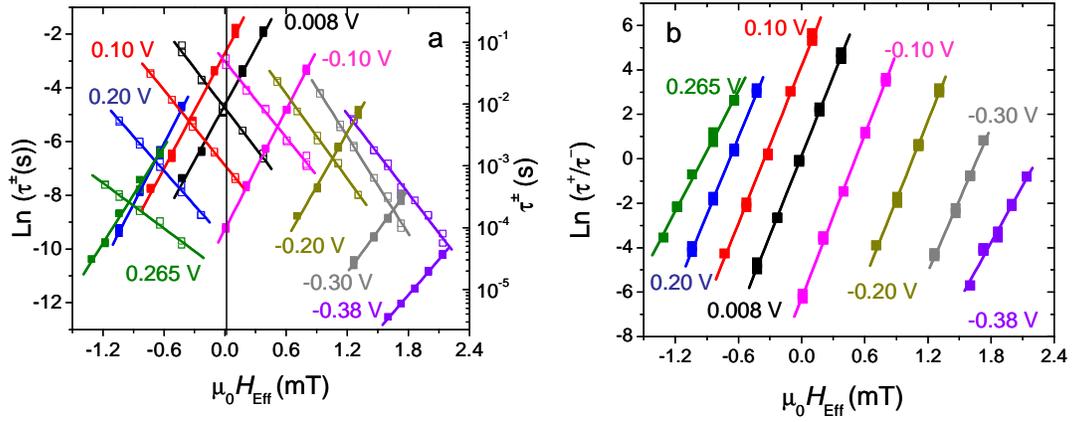

FIG 3 (a) Summary of the measured relaxation times at $T = 303$ K for both $\tau^+$ (the P-to-AP (■) transition) and $\tau^-$ (the AP-to-P (□) transition) as a function of $H_{Eff}$ for several different applied voltages, as indicated in the plot, along with linear fits. (b) $d(\mathrm{Ln}(\tau^+/\tau^-))/d(\mu_0 H_{Eff})$ for the same applied voltages along with linear fits.

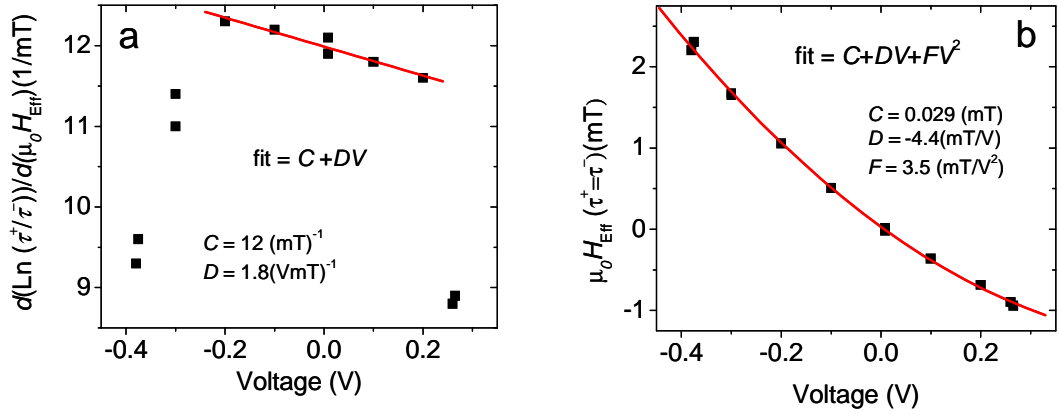

FIG 4 (a) Measured slopes of the data in Fig. 3(b), $d(Ln(\tau^+/\tau^-))/d(\mu_0 H_{Eff})$ vs. applied voltage along with a linear fit for $|V| < 0.2$ V. (b) The measured value of $H_{Eff}$-intercept of the data in Fig. 3(b) along with a quadratic fit. The error bars in determining $\mu_0 H_{Eff}(\tau^+=\tau^-)$ are $\pm 0.025$ mT and are smaller than the data points.